\documentclass[review,12pt]{elsarticle}

\usepackage[a4paper,margin=1in]{geometry}
\usepackage{amsmath,amssymb,bm}
\usepackage{physics}
\usepackage{graphicx}
\usepackage{subcaption}
\usepackage{xspace}
\usepackage{xcolor}
\usepackage{booktabs}
\usepackage{fontspec}
\graphicspath{{Figs/}}
\usepackage{amsmath}
\usepackage{float}

\newcommand{\includegraphicsmaybe}[2][]{%
	\IfFileExists{#2.pdf}{\includegraphics[#1]{#2.pdf}}{%
		\IfFileExists{#2.png}{\includegraphics[#1]{#2.png}}{%
			\IfFileExists{#2.jpg}{\includegraphics[#1]{#2.jpg}}{%
				\fbox{\textsf{Missing figure: #2}}%
			}%
		}%
	}%
}

\newcommand{\kB}{k_{\mathrm B}}
\newcommand{\e}{\mathrm e}
\newcommand{\I}{\mathrm I} 

\newcommand{\MFTI}{MFTI\xspace}      
\newcommand{\TSCPI}{TSCPI\xspace}    
\newcommand{\MDMSD}{MDMSD\xspace}    

\journal{Nature Communications}

\begin{document}
	
\begin{frontmatter}
\title{Dual Role of Mobile Interstitials in Defect Kinetics: From Retardation to Acceleration}

\author[address1]{Shihao Zhang\fnref{myfootnote}}
\author[address1]{Shihao Zhu\fnref{myfootnote}}
\fntext[myfootnote]{These authors contributed equally to this work.}

\author[address1]{Junping Du}
\author[address1]{Shuhei Shinzato}
\author[address2]{Ju Li\corref{correspondingauthor}}
\cortext[correspondingauthor]{Corresponding author}
\ead{liju@mit.edu}
\author[address1]{Shigenobu Ogata\corref{correspondingauthor}}
\ead{ogata.shigenobu.es@osaka-u.ac.jp}

\address[address1]{Department of Mechanical Science and Bioengineering, Graduate School of Engineering Science, The University of Osaka, Osaka, 560-8531, Japan}
\address[address2]{Department of Nuclear Science and Engineering and Department of Materials Science and Engineering, Massachusetts Institute of Technology, Cambridge, Massachusetts 02139, USA}

\begin{abstract}
Mobile interstitial atoms redistribute while defects migrate, complicating prediction of defect kinetics and leading to contradictory reports of acceleration and retardation. We formulate defect activation in a grand-canonical ensemble at fixed interstitial chemical potential and define a pathway free-energy landscape $\Delta G(\Lambda;T,\mu)$ and the corresponding activation free energy $\Delta G^{\ddagger}$. We compute these quantities using both hyperplane-constrained thermodynamic integration via a mean force relation and a two-state chemical-potential integration, denoted \TSCPI. The latter requires sampling only the initial and transition states; a single chemical-potential integration then yields $\Delta G^{\ddagger}(T,\mu)$ across a wide $\mu$ range, enabling rapid mapping over temperature and chemical potential. Applied to vacancy diffusion in FCC, BCC, and HCP metals with H (including in plane and cross plane diffusion in Zr) and in BCC W with He, both free energy routes agree with diffusion coefficients from molecular dynamics mean square displacements. The resulting maps reveal regimes of suppression and enhancement, including a crossover from suppression to enhancement with increasing hydrogen concentration. A site occupancy analysis links barrier shifts to state dependent site spectrum changes and transferable interstitial interaction terms.
\end{abstract}


\end{frontmatter}

%
%
%
%
%



\newpage
\section*{Introduction}

Hydrogen and helium are among the most mobile interstitial species in metals and are widely present in service environments, introduced by corrosion related processes or generated by nuclear transmutation reactions \cite{dwivedi2018hydrogen,robertson2015hydrogen,gilbert2013neutron,gilbert2012integrated}. They can severely degrade mechanical performance, manifesting as hydrogen embrittlement and helium embrittlement \cite{dwivedi2018hydrogen,robertson2015hydrogen,Li2020,Lynch2012,Baskes1986}. These phenomena are tightly linked to the evolution of lattice defects such as vacancies, dislocations, and grain boundaries, which also act as efficient trapping and redistribution sites for mobile interstitials \cite{chen2025hydrogen,Cao2024,Wu2025,yu2024hydrogen}. Quantifying how mobile interstitials modify defect kinetics is therefore essential for predictive models of degradation.

Despite its importance, a unified description remains elusive. Reported interstitial effects on defect kinetics are not even qualitatively consistent: in some materials and conditions, interstitial charging accelerates defect processes, whereas in others it retards them. Even for the simplest thermally activated process, vacancy diffusion, contradictory predictions have been reported for hydrogen-charged metals, including diffusion enhancement \cite{Du2020} and diffusion suppression \cite{Liu2019,Hachet2022a}, with no consensus framework to reconcile the trends across materials, temperatures, and charging conditions. For vacancy diffusion in particular, many studies have interpreted hydrogen effects in terms of vacancy and hydrogen complexes with a prescribed trapped occupancy and have discussed limiting scenarios ranging from concerted migration of an intact complex to dissociation-assisted pathways in which hydrogen transiently detaches and retraps during vacancy motion \cite{Hayward2013,Mitsuhara2023,Huang2022}. While such pictures are useful in specific limits, they also highlight that the interstitial population in and around a migrating defect is not fixed under realistic charging conditions controlled by an external reservoir.

A key reason is that a kinetic observable is not defined by a single activated event with a single interstitial arrangement. Rather, quantities such as a vacancy diffusion coefficient or a dislocation velocity are ensemble averaged over many barrier crossing events, each occurring under a different instantaneous interstitial configuration. When interstitials are sufficiently mobile, these configurations are sampled under conditions controlled by a reservoir variable such as chemical potential, while the defect repeatedly migrates. Accordingly, the relevant activation free energy should be defined for an ensemble of interstitial configurations sampled at fixed $N_{\mathrm{lat}}$, volume $V$, temperature $T$, and $\mu$, rather than for a single frozen arrangement. Here $N_{\mathrm{lat}}$ denotes the number of host lattice atoms in the simulation cell, and $\mu$ denotes the chemical potential of the mobile interstitial species $\mathrm I$ imposed by an external reservoir. This perspective also implies that interstitial charging does not necessarily affect defect mobility monotonically and can produce crossovers between suppression and enhancement as temperature and charging vary. Overall, mobile interstitial atoms such as hydrogen and helium can strongly modulate defect kinetics, yet a unified thermodynamic description has been lacking because interstitials redistribute while defects migrate.

In the present work, we formulate defect activation free energies at fixed interstitial chemical potential within a grand canonical framework. Hyperplane constrained thermodynamic integration provides the free energy profile along a reference reaction pathway and yields the activation free energy at the transition state. A computationally efficient two state chemical potential integration, \TSCPI, expresses barrier changes through the difference in interstitial uptake between the initial and transition states and enables wide mapping over temperature and chemical potential. Using vacancy diffusion as a benchmark, we validate the framework against direct molecular dynamics estimates across H-charged FCC, BCC, and HCP metals (including in-plane and cross-plane diffusion in HCP Zr) and He-charged BCC W. Wide-range maps reveal regimes of diffusion suppression and enhancement, including crossovers between them as hydrogen concentration increases. Effective site-energy spectra and cumulative site counts, augmented by an effective H-H interaction that is required at higher occupancy, provide a unified qualitative explanation for the crossover behavior in Cu and Pd and the monotonic suppression in Fe (BCC).

\section*{Results}

\subsection*{\textbf{Grand canonical activation free energies under mobile interstitials}}

\begin{figure}[H]
	\centering
	\includegraphics[width=0.6\textwidth]{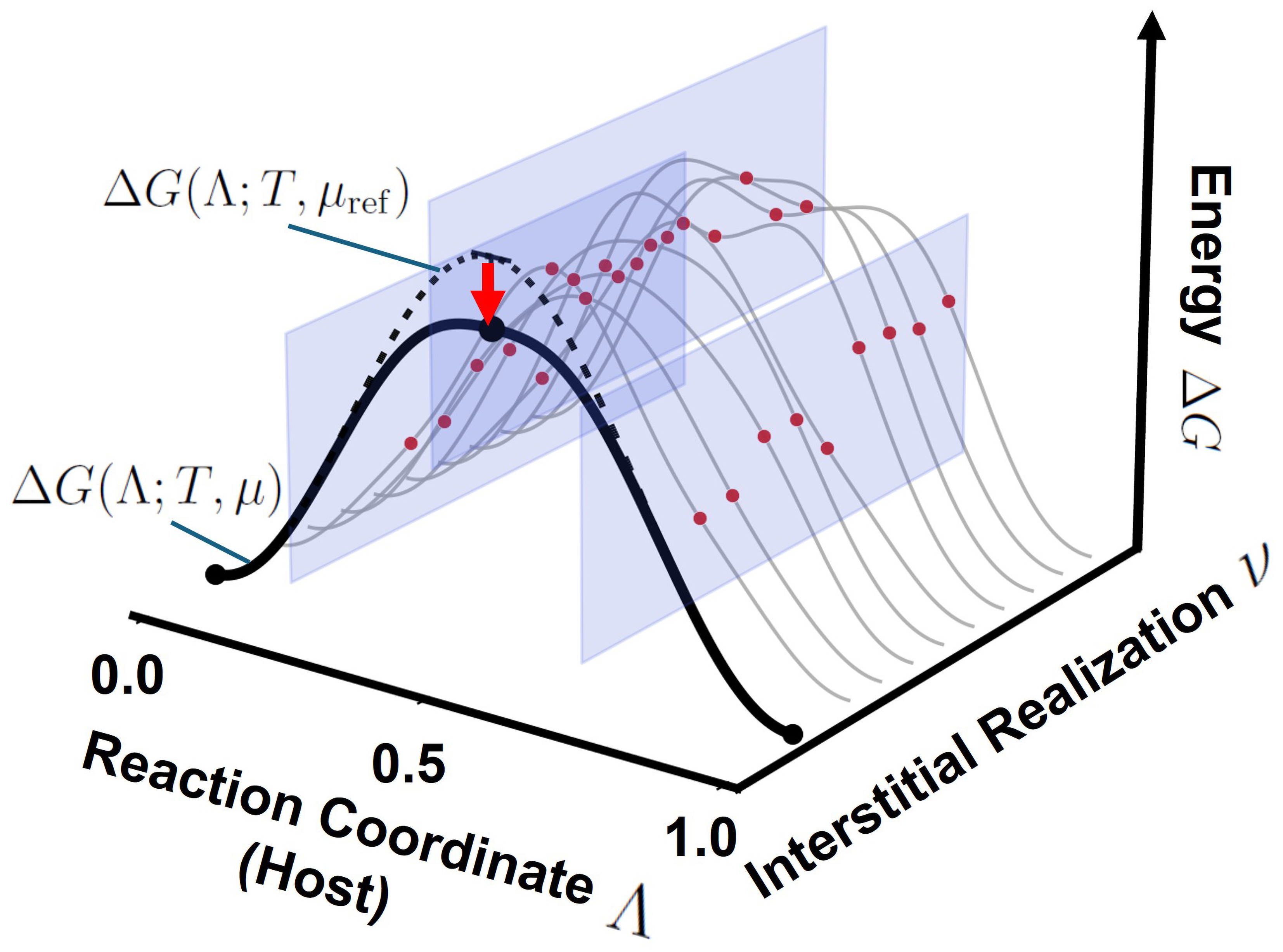}
	\caption{\textbf{Schematic of grand canonical activation free energies under mobile interstitials.} A reference defect migration pathway is parameterized by $\Lambda$. The solid and dotted curves represent $\Delta G(\Lambda;T,\mu)$ and $\Delta G(\Lambda;T,\mu_{\rm ref})$, respectively; both can be obtained from thermodynamic integration of the constrained mean force [Eqs.~\eqref{eq:DG_global_concat} and \eqref{eq:TI_lambda_segment}]. At a given $\Lambda$, that is, for a given host lattice configuration along the pathway, the difference $\Delta G(\Lambda;T,\mu)-\Delta G(\Lambda;T,\mu_{\rm ref})$ can alternatively be evaluated from the chemical-potential integration formula [Eqs.~\eqref{eq:DG_global_concat} and \eqref{eq:DG_muTI_general_ref}]. In the present schematic this difference is negative, but it can also be positive depending on the host-interstitial system and thermodynamic conditions.}
	\label{fig:schematic_TI}
\end{figure}

Here we develop a unified grand-canonical activation-free-energy theory for defect kinetics under mobile interstitials. A crystalline host with fixed $N_{\mathrm{lat}}$, $V$, and $T$ is coupled to a reservoir with chemical potential $\mu$, allowing the number of interstitial atoms of species $\mathrm I$ to fluctuate. For each segment $k$ and hyperplane label $\lambda$ along a reference reaction pathway, the hyperplane-constrained grand partition function is constructed by restricting the host configuration to piecewise linear hyperplanes $\mathcal H_{k,\lambda}$ (see Appendixes A-C):
\begin{flalign}
	& \Xi_{\mathcal H}^{(k)}(\mu,V,T;\lambda) = \nonumber \\
	& \sum_{N=0}^{\infty}
	\frac{z^N}{N!\,h^{3(N_{\mathrm{lat}}+N)}}
	\int \dd\mathbf X\,\dd\mathbf P
	\int \prod_{i=1}^{N}\dd\mathbf x_i\,\dd\mathbf p_i\;
	\delta\!\Big(\hat{\mathbf n}_k\cdot(\mathbf X-\mathbf X^{(k)})-\lambda L_k\Big)\,
	\e^{-\beta H(\mathbf X,\mathbf P,\{\mathbf x_i,\mathbf p_i\};N)}.
	&&
	\label{eq:Xi_hyperplane_segment}
\end{flalign}
The corresponding constrained grand potential is
\begin{align}
	\Omega_{\mathcal H}^{(k)}(\lambda;\mu,T)\equiv -\kB T\ln \Xi_{\mathcal H}^{(k)}(\mu,V,T;\lambda).
	\label{eq:Omega_hyperplane_segment}
\end{align}
We define the free energy change along segment $k$ relative to its initial hyperplane:
\begin{align}
	\Delta G_k(\lambda;T,\mu)\equiv
	\Omega_{\mathcal H}^{(k)}(\lambda;\mu,T)-\Omega_{\mathcal H}^{(k)}(0;\mu,T).
	\label{eq:DG_segment_def}
\end{align}
Eqs. \eqref{eq:Xi_hyperplane_segment}-\eqref{eq:DG_segment_def} can be viewed as an equilibrium reduction of a path ensemble, in which repeated activated events sample different instantaneous interstitial configurations at fixed $\mu$ (see Fig. \ref{fig:schematic_TI} and Appendix D). In the piecewise linear representation, the cumulative free energy profile along the pathway is evaluated by concatenation:
for $s=\Lambda S_{\mathrm{tot}}\in[S_k,S_{k+1}]$,
\begin{align}
	\Delta G(\Lambda;T,\mu)
	\approx
	\sum_{m=0}^{k-1}\Delta G_m(1;T,\mu)
	+
	\Delta G_k(\lambda_k(\Lambda);T,\mu).
	\label{eq:DG_global_concat}
\end{align}

Because $\lambda$ enters through the hyperplane constraint, the derivative of $\Omega_{\mathcal H}^{(k)}$ is expressed as a mean force. For the piecewise linear hyperplanes $\mathcal H_{k,\lambda}$, one obtains
\begin{align}
	\dv{\Omega_{\mathcal H}^{(k)}}{\lambda}
	=
	-L_k\left\langle \hat{\mathbf n}_k\cdot\mathbf F_{\mathbf X}\right\rangle_{\mu VT,k,\lambda},
	\ 
	\mathbf F_{\mathbf X}\equiv -\nabla_{\mathbf X}U,
	\label{eq:dOmega_dlambda_force}
\end{align}
where $U$ is the potential energy and $\langle\cdots\rangle_{\mu VT,k,\lambda}$ denotes the hyperplane constrained grand canonical average on segment $k$.
Integrating Eq. \eqref{eq:dOmega_dlambda_force} gives
\begin{eqnarray}
	\Delta G_k(\lambda;T,\mu)
	&=&
	\int_{0}^{\lambda}\dv{\Omega_{\mathcal H}^{(k)}}{\lambda'}\,\dd\lambda' \nonumber \\
	&=&
	-L_k\int_{0}^{\lambda}
	\left\langle \hat{\mathbf n}_k\cdot\mathbf F_{\mathbf X}\right\rangle_{\mu VT,k,\lambda'}\dd\lambda'.
	\label{eq:TI_lambda_segment}
\end{eqnarray}

Furthermore, we make the $\mu$ dependence explicit. A grand canonical identity gives, for each constrained state,
\begin{align}
	\pdv{\Omega_{\mathcal H}^{(k)}(\lambda;\mu)}{\mu}
	=
	-\langle N\rangle_{\mu VT,k,\lambda},
	\label{eq:dOmega_dmu}
\end{align}
where $\langle N\rangle_{\mu VT,k,\lambda}$ is the mean number of interstitial atoms in a volume $V$ at temperature $T$ under the hyperplane constraint on segment $k$ at $\lambda$.
Integrating Eq. \eqref{eq:dOmega_dmu} from a reference $\mu_{\mathrm{ref}}$ yields
\begin{align}
	\Omega_{\mathcal H}^{(k)}(\lambda;\mu)
	=
	\Omega_{\mathcal H}^{(k)}(\lambda;\mu_{\mathrm{ref}})
	-
	\int_{\mu_{\mathrm{ref}}}^{\mu}
	\langle N\rangle_{\mu' VT,k,\lambda}\dd\mu'.
	\label{eq:Omega_muTI_ref}
\end{align}
Taking the difference between $\lambda$ and $0$ gives a general decomposition of the segment free energy profile:
\begin{eqnarray}
	&&\Delta G_k(\lambda;T,\mu)
	=
	\Delta G_k(\lambda;T,\mu_{\mathrm{ref}}) \nonumber \\
	&&-
	\int_{\mu_{\mathrm{ref}}}^{\mu}
	\Big[
	\langle N\rangle_{\mu' VT,k,\lambda}
	-
	\langle N\rangle_{\mu' VT,k,0}
	\Big]\dd\mu'.
	\label{eq:DG_muTI_general_ref}
\end{eqnarray}

If the reference pathway contains a transition state located on segment $k^{\ddagger}$ at $\lambda=\lambda^{\ddagger}$, the grand canonical activation free energy is given by $\Delta G^{\ddagger}(T,\mu)\equiv \Delta G_{k^{\ddagger}}(\lambda^{\ddagger};T,\mu)$, referred to as the \MFTI method (mean force thermodynamic integration along the constrained pathway, see Appendix E). Alternatively, if the transition-state hyperplane is weakly affected by interstitial charging, its location $(k^{\ddagger},\lambda^{\ddagger})$ can be fixed from an uncharged reference pathway. Evaluating Eq.~\eqref{eq:DG_muTI_general_ref} at this fixed transition state also yields $\Delta G^{\ddagger}(T,\mu)$, referred to as the \TSCPI method, i.e., two-state chemical potential integration using only the initial and transition states.

Under fixed $T$ and $\mu$, the diffusion coefficient can be written in Arrhenius form as 
\begin{align}
	D(T,\mu)=D_0\exp\!\left[-\frac{\Delta G^{\ddagger}(T,\mu)}{\kB T}\right],
	\label{eq:arrhenius_D}
\end{align}
where $D_0$ is a temperature independent kinetic prefactor for a given host and jump geometry. If the prefactor is only weakly affected by interstitial charging, the diffusion enhancement and suppression factor reduces to a barrier only expression:
\begin{align}
	\frac{D(T,\mu)}{D(T,\mu_{\rm ref})}
	\approx
	\exp\!\left[
	-\frac{\Delta G^{\ddagger}(T,\mu)-\Delta G^{\ddagger}(T,\mu_{\rm ref})}{\kB T}
	\right].
	\label{eq:ratio_from_DG}
\end{align}
Equations~\eqref{eq:DG_muTI_general_ref} and \eqref{eq:ratio_from_DG} establish a direct thermodynamic connection between interstitial statistics and defect kinetics, and enables efficient evaluation of activation free energies over wide ranges of temperature and chemical potential using only the initial and transition states.

\subsection*{\textbf{Diffusion enhancement and suppression as functions of temperature $T$ and concentration $c_{\mathrm H}^{\mathrm{bulk}}$}}

To validate the theory in a quantitatively testable setting, we focus on vacancy diffusion as a prototypical thermally activated defect process. We compare three routes: \TSCPI, \MFTI, and \MDMSD, the last of which estimates diffusion directly from molecular-dynamics trajectories via the mean-square displacement. We examine hydrogen in FCC Cu, Ni, and Pd, BCC Fe, and HCP Zr, as well as helium in BCC W; for Zr-H, both in-plane and cross-plane diffusion are analyzed. For each system, diffusion enhancement and suppression factors relative to the uncharged reference, evaluated using all three methods, are in excellent agreement, as shown in Fig.~\ref{fig:diffusion_ratios}; the corresponding results for Ni and Zr with H, and W with He are provided in Supplementary Figs.~S1 and S2. Because both \MFTI and \TSCPI convert the interstitial induced change in activation free energy into a diffusion ratio by assuming a weak interstitial dependence of the kinetic prefactor, their agreement with direct \MDMSD results, which do not rely on this assumption, indicates that prefactor variations with interstitial charging are minor in the present regimes. This consistency supports the central premise that the dominant effect of mobile interstitials on vacancy diffusion is captured by the grand-canonical renormalization of the activation free energy.

Beyond validation, \TSCPI enables efficient computation of activation-free-energy changes and construction of three-dimensional maps of diffusion enhancement and suppression as functions of temperature $T$ and concentration $c_{\mathrm H}^{\mathrm{bulk}}$ (see Fig.~\ref{fig:diffusion_ratios} and Supplementary Figs.~S1 and S2). The FCC Cu-H, Pd-H, and Ni-H systems exhibit clear crossovers: vacancy diffusion is mildly retarded at low $c_{\mathrm I}^{\mathrm{bulk}}$ and low temperature, but becomes accelerated beyond a threshold. Within our framework, these crossovers correspond to sign changes in the barrier shift $\Delta\Delta G^{\ddagger}(T,\mu)\equiv \Delta G^{\ddagger}(T,\mu)-\Delta G^{\ddagger}(T,\mu_{\rm ref})$, reflecting that interstitials stabilize the initial and transition states differently as $\mu$ (and hence $c_{\mathrm I}^{\mathrm{bulk}}$) and temperature increase.

\begin{figure}[H]
	\centering
	\includegraphics[width=1.0\textwidth]{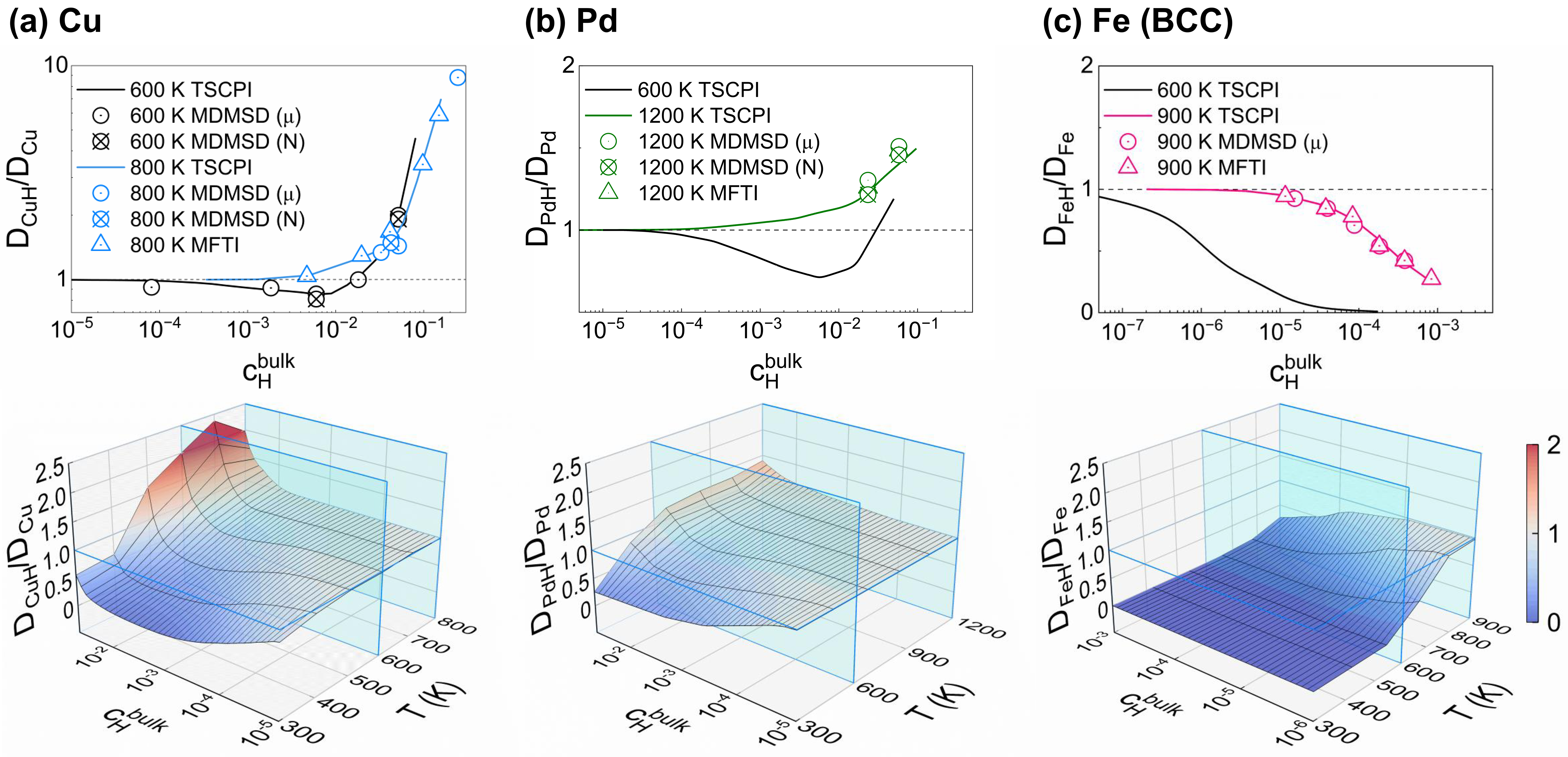}
	\caption{\textbf{Diffusion enhancement and suppression as functions of temperature $T$ and concentration $c_{\mathrm H}^{\mathrm{bulk}}$.} (a) Cu, (b) Pd, and (c) Fe (BCC): in each material, the upper panel shows the diffusion enhancement or suppression factor extracted along a constant-temperature slice to make its variation with concentration $c_{\mathrm H}^{\mathrm{bulk}}$ clear, while the lower panel shows the corresponding three-dimensional map computed using \TSCPI\ as a function of $T$ and $c_{\mathrm H}^{\mathrm{bulk}}$ (equivalently controlled by $\mu_{\mathrm H}$). The constant-temperature plane corresponding to the upper panel is indicated in each lower map. Symbols and lines in the upper panels compare \TSCPI, \MFTI, and \MDMSD; the latter was performed under conditions of either a fixed hydrogen chemical potential $\mu$ or a fixed number of hydrogen atoms $N$ in the simulation cell. The close agreement among the three routes validates the grand-canonical activation-free-energy description across materials, temperatures, and charging conditions.}
	\label{fig:diffusion_ratios}
\end{figure}

Vacancy diffusion is central to many thermally activated processes in metals, including dislocation climb \cite{HirthLothe1982}, precipitate coarsening \cite{Lifshitz1961}, void evolution \cite{BrailsfordBullough1972}, radiation-induced segregation \cite{Wiedersich1979}, irradiation swelling \cite{BrailsfordBullough1972}, and the annealing and shrinkage of irradiation-induced dislocation loops and voids \cite{EvansEldrup1975,Yuan2022}. In most cases, the kinetics depend on both the vacancy concentration and the mobility of individual vacancies. Therefore, a measured rate change cannot be assigned uniquely to a change in single-vacancy mobility. Under non-equilibrium or externally constrained conditions, however, the vacancy population is set by the initial state or by the imposed environment. Examples include the recovery of quenched-in vacancies \cite{Koehler1957,Elsayed2023}, the annealing of irradiation-produced vacancies \cite{EvansEldrup1975,Hautojarvi1985}, and ageing under nearly fixed vacancy supersaturation \cite{Hachet2022}. These processes are therefore more sensitive to the mobility of pre-existing vacancies and provide a more direct route to examine how hydrogen modifies vacancy migration.
	
Available mobility-sensitive experiments indicate that hydrogen retards vacancy motion at low temperature, broadly consistent with the low-temperature trend shown in Fig.~\ref{fig:diffusion_ratios}. In high-purity Al, positron-annihilation spectroscopy shows that hydrogen delays monovacancy recovery from $\sim$220 K to about $\sim$280 K, indicating vacancy immobilization through vacancy--hydrogen pair formation \cite{Elsayed2023}. In electron-irradiated Nb and Ta, positron-lifetime measurements show that hydrogen decoration of irradiation-induced vacancies shifts the vacancy-clustering recovery stage from about 220--260 K in pure samples to about 380--450 K in hydrogen-charged samples, showing that vacancy--hydrogen complex formation retards the effective mobility of pre-existing vacancies \cite{Hautojarvi1985}. In Al--Mg--Si alloys aged at approximately constant vacancy concentration, hydrogen slows precipitate coarsening and delays softening while leaving the precipitate volume fraction and composition nearly unchanged. Classical coarsening analysis attributes this behavior to an increase in the effective vacancy migration energy by about 5\%, when the hydrogen concentration is close to the vacancy concentration \cite{Hachet2022}.
	
Direct experimental evidence for hydrogen-enhanced single-vacancy mobility under high-$c_{\mathrm H}^{\mathrm{bulk}}$ and high-temperature conditions is still lacking. Further experiments that independently control or measure vacancy concentration are desirable. It should be noted that many reports of hydrogen-enhanced vacancy-mediated kinetics mainly reflect a hydrogen-induced increase in vacancy equilibrium concentration, as in the superabundant-vacancy effect \cite{Fukai1993,Fukai2003}. Such experiments demonstrate accelerated macroscopic kinetics, but they do not prove that a hydrogenated vacancy has a lower migration barrier than a bare vacancy.

\subsection*{\textbf{Site occupancy picture and effective site-energy distributions}}

The \TSCPI method identifies the key thermodynamic control parameter for barrier modulation as the difference in grand-canonical uptake between the initial and transition states. To provide a microscopic interpretation, we analyze the site occupancy picture for the representative Cu, Pd, and Fe (BCC) with H systems, focusing on the site-occupancy free energy levels and their degeneracies in the initial and transition states and the role of H-H interactions beyond the independent site limit.

For each constrained host state (initial and transition state), we enumerate a set of candidate interstitial sites in the vicinity of the migrating vacancy and the transition state environment within a prescribed cutoff region that captures the dominant binding landscape.
We then group symmetry equivalent sites into classes $m=1,2,\ldots$ and characterize each class by a site-occupancy free energy level $\phi_m(T)$ and a degeneracy $g_m$, where $g_m$ counts the number of sites belonging to the class.

\begin{figure}[H]
	\centering
	\includegraphics[width=0.6\textwidth]{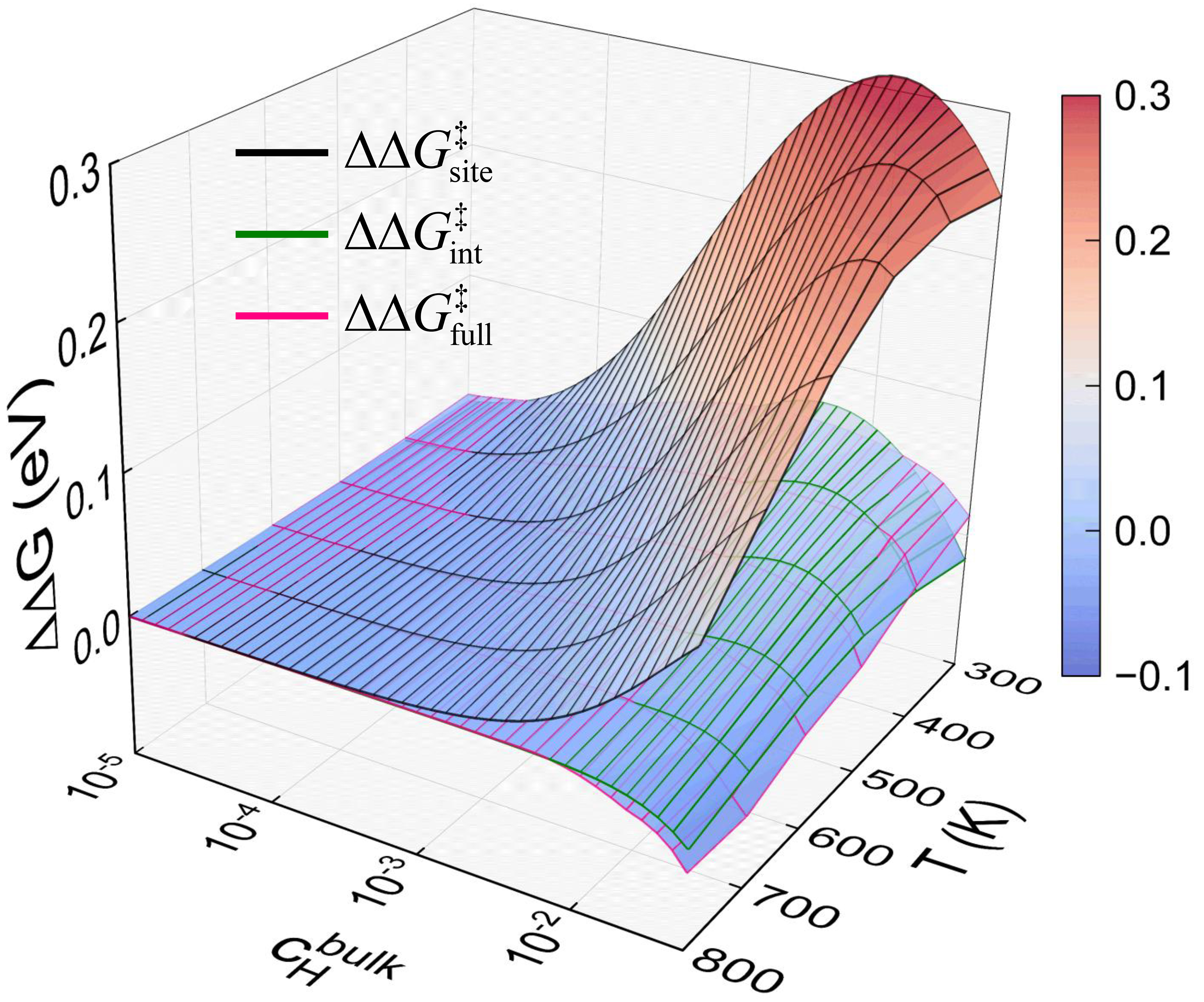}
	\caption{\textbf{Role of H-H interactions in the barrier shift in Cu.} Charging-induced changes in the activation free energy calculated using the independent-site model based on the noninteracting site spectra $\phi_j$ in Supplementary Fig.~S3, $\Delta\Delta G^{\ddagger}{\mathrm{site}}(T,\mu)$, and from the effective site energies $\phi_j^{\mathrm{HH}}$ obtained by fitting $V_j$ (Supplementary Fig.~S4), $\Delta\Delta G^{\ddagger}{\mathrm{int}}(T,\mu)$, are compared with the full \TSCPI results, $\Delta\Delta G^{\ddagger}_{\mathrm{full}}(T,\mu)$. Neglecting H-H interactions leads to substantial errors, whereas incorporating them via effective site energies yields quantitative agreement with the full grand-canonical result. Analogous interaction corrections for Pd and Fe (BCC) are shown in Supplementary Fig.~S5.}
	\label{fig:HH_interaction_fit}
\end{figure}

Given the site spectra $(\phi_m,g_m)$ for a constrained state, the independent site lattice gas model yields
\begin{align}
	\Omega_{\mathrm{site}}(\mu,T)
	=
	-\kB T \sum_m g_m
	\ln\!\left[1+\exp\!\big(\beta(\mu-\phi_m(T))\big)\right],
	\label{eq:Omega_site_degeneracy}
\end{align}
and
\begin{align}
	\langle N\rangle_{\mathrm{site}}(\mu,T)
	=
	\sum_m g_m
	\frac{1}{1+\exp\!\big(\beta(\phi_m(T)-\mu)\big)}.
	\label{eq:N_site_degeneracy}
\end{align}
Using these expressions in the noninteracting version of our chemical potential decomposition, we compute the charging induced free energy changes for the initial and transition states and hence the predicted barrier shift without H-H interactions,
\begin{align}
	\Delta\Delta G^{\ddagger}_{\mathrm{site}}(T,\mu)
	\equiv
	\Big[\Delta G^{\ddagger}(T,\mu)-\Delta G^{\ddagger}(T,\mu_{\mathrm{ref}})\Big]_{\mathrm{site}} .
	\label{eq:DDG_site_def}
\end{align}
Comparing $\Delta\Delta G^{\ddagger}_{\mathrm{site}}(T,\mu)$ with the fully resolved barrier shifts obtained from \TSCPI (and validated by \MFTI and \MDMSD) isolates the contribution that is not captured by independent occupancy of the site spectrum. For simplicity, in the analysis below we approximate the site energy levels $\phi_m(T)$ by their $T=0$~K values obtained from relaxed structures and treat them as temperature independent. As shown in Fig.~\ref{fig:HH_interaction_fit}, neglecting H-H interactions can lead to a large deviation from the full barrier shift, particularly at higher chemical potentials where multiple H atoms occupy the defect environment.

We attribute the residual difference between the full grand canonical result and the noninteracting site prediction to H-H interactions, $\Delta\Delta G^{\ddagger}_{\mathrm{full}}(T,\mu) - \Delta\Delta G^{\ddagger}_{\mathrm{site}}(T,\mu)$, and represent it by an effective correction to the site-occupancy free energy levels,
\begin{align}
	\phi_j^{\mathrm{HH}}(T) = \phi_j(T) + V_j,
	\label{eq:phi_HH_def}
\end{align}
where $V_j$ captures the net energetic shift associated with H-H interactions for occupancy of site $j$ under the relevant constrained host state. The fitted $V_j$ spectra and the resulting $\phi_j^{\mathrm{HH}}$ distributions are reported in Supplementary Fig.~S4. Here $\phi_j$ and $\phi_j^{\mathrm{HH}}$ are evaluated in the $T=0$~K approximation, so $V_j$ is treated as temperature independent.
Using $\phi_j^{\mathrm{HH}}(T)$ in the site occupancy model yields an interaction corrected estimate of $\Delta\Delta G^{\ddagger}(T,\mu)$, denoted as $\Delta\Delta G^{\ddagger}_{\mathrm{int}}(T,\mu)$.
As shown in Fig.~\ref{fig:HH_interaction_fit}, including H-H interactions in this way enables quantitative reproduction of the full barrier shifts over the investigated chemical potential range.

\begin{figure}[H]
	\centering
	\includegraphics[width=1.0\textwidth]{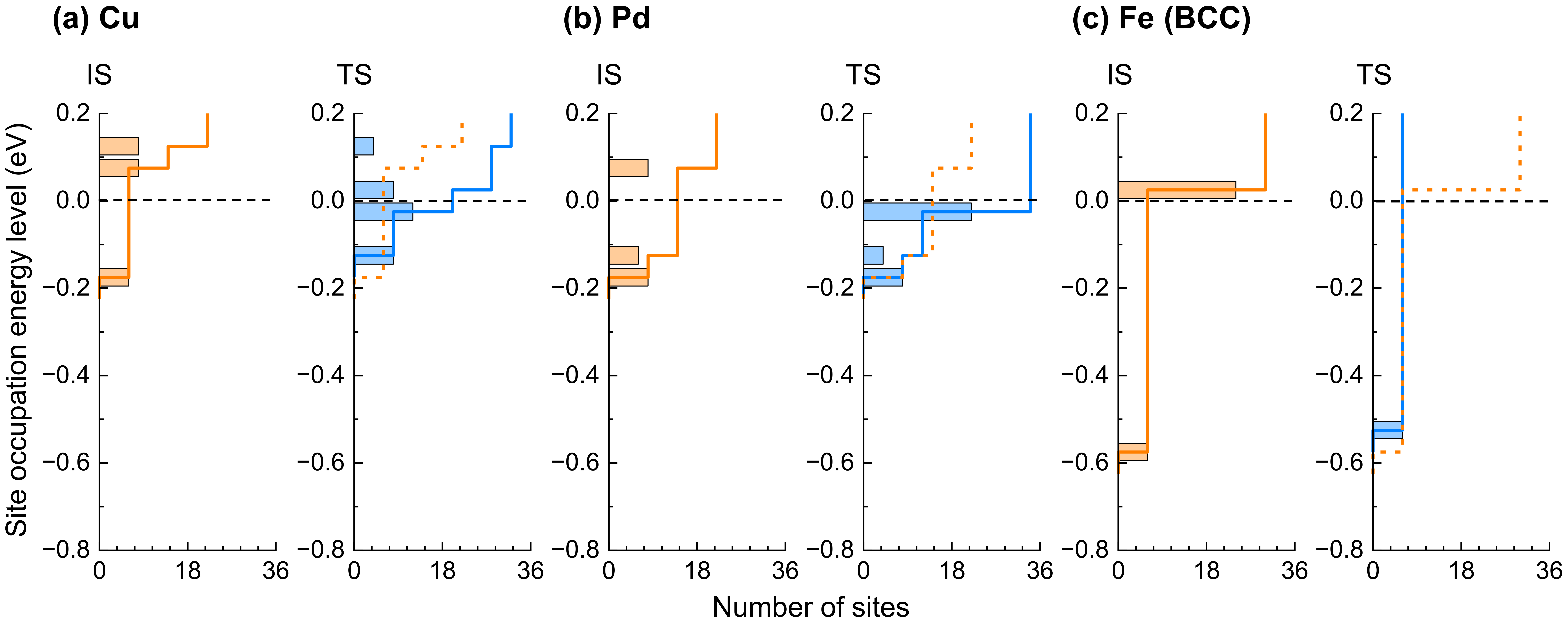}
	\caption{\textbf{Effective site-energy distributions that include H-H interactions.} Site energies are referenced to the bulk octahedral site in Cu and Pd, and to the bulk tetrahedral site in Fe (BCC). Panels show (a) Cu with H, (b) Pd with H, and (c) Fe (BCC) with H for the initial and transition states. The effective site energies $\phi_j^{\mathrm{HH}}$ are shown as a spectrum versus the site index (number of sites), and the cumulative number of sites from $-\infty$ up to a given site energy is overlaid as a solid curve. In the transition-state panels, the cumulative curve of the initial state is additionally reproduced as a dashed curve to highlight the difference between the two states and the resulting crossover or lack thereof.}
	\label{fig:phiHH_spectra}
\end{figure}

Figure~\ref{fig:phiHH_spectra} summarizes the effective site-energy distributions $\phi_j^{\mathrm{HH}}$ and the corresponding cumulative number of available sites for the initial and transition states in Cu, Pd, and Fe (BCC) with H. In Cu and Pd, the cumulative site count of the initial state exceeds that of the transition state only in the deepest-energy tail, whereas at higher energies the transition state accumulates substantially more sites. As Eq.~\eqref{eq:DG_muTI_general_ref} shows, this change in relative uptake yields a positive barrier shift at low $\mu$ and a negative barrier shift at higher $\mu$, producing the observed crossover from diffusion suppression to enhancement with increasing bulk hydrogen concentration. By contrast, in Fe (BCC) the initial-state cumulative count is comparable to or greater than that of the transition state over nearly the entire spectrum, so the initial state consistently traps more H and the barrier increases over the full range of $\mu$, leading to diffusion suppression. These spectrum-based trends fully rationalize the material-dependent responses in Fig.~\ref{fig:diffusion_ratios}. 

\section*{Method}

\subsection*{\textbf{Details of simulation settings}}

All atomistic simulations were performed with the Large-scale Atomic/Molecular Massively Parallel Simulator (LAMMPS) \cite{Thompson2022}. Equilibrium interstitial configurations at each constrained host state were sampled by hybrid MD and grand canonical Monte Carlo (GCMC), alternating NVT molecular dynamics updates of the host with $\mu$VT insertion and deletion moves for the interstitial species to maintain the prescribed chemical potential \cite{Frenkel2002}. The MD time step was 0.5 fs. Every 10 MD steps, 100 GCMC insertion/deletion trials of H or He were attempted, allowing the supercell composition to relax toward the prescribed solution chemical potential, \(\mu_{\mathrm{I}}\).

We employed our first-principles–trained machine-learning interatomic potentials (MLIPs) for Pd-H, Fe–H \cite{Zhang2024}, and Ni–H. In particular, the Pd-H and Ni–H MLIPs were specifically trained to describe vacancy–H interactions within the framework of the moment tensor potential formalism \cite{Shapeev2016,Novoselov2019}. Both MLIPs accurately reproduce DFT-level potential-energy surfaces and atomic forces while maintaining high computational efficiency. Embedded-atom-method (EAM) potentials were used for Cu–H \cite{Du2020}, Zr–H \cite{Wimmer2020}, and W–He \cite{Bonny2014}. For Cu–H, we combine the Cu EAM potential developed by Mishin \textit{et al.} \cite{Mishin2001} with an H potential and a Cu–H cross interaction fitted in our previous work \cite{Du2020}. The resulting H–vacancy interaction in Cu has been extensively validated; see Sec.~S9 of the Supplemental Material of Ref.~\cite{Du2020}.

\subsection*{\textbf{Interstitial uptake $\langle N \rangle_{\mu VT}$ at initial and transition states}}

The interstitial uptake $\langle N \rangle_{\mu VT}$ used in the TSCPI method (Eq. (9)) was determined as $\langle N^{\rm supercell}_{\rm I} \rangle - N^{\rm supercell}_{\rm M} c^{\rm bulk}_{\rm I}$, where $N^{\rm supercell}_{\rm I}$ is the total number of interstitial atoms in the supercell containing a monovacancy in the initial or transition states, $N^{\rm supercell}_{\rm M}$ is the number of metal atoms in the same supercell, and $c^{\rm bulk}_{\rm I}$ is the bulk interstitial atomic concentration corresponding to the specified chemical potential $\mu$. The values of $\langle N^{\rm supercell}_{\rm I} \rangle$ and $c^{\rm bulk}_{\rm I}$ were calculated using the hybrid MD/GCMC method at different temperatures $T$ and interstitial chemical potential $\mu$. Angle brackets \(\langle\rangle\) indicate ensemble averages over MD/GCMC configurations at fixed \(\mu\) and \(T\). Supercells contained 216 atoms for Fe, 1024 atoms for Cu, Pd, and Ni, 1440 atoms for Zr, and 720 atoms for W. A single vacancy was created by removing one metal atom from the perfect lattice. Periodic boundary conditions were applied in all three Cartesian directions. Ensemble averages were computed over \(1.2\times10^{6}\) MD steps following a \(3.0\times10^{5}\)-step equilibration.

\subsection*{\textbf{Mean force $\mathbf{F_X}$ worked on the migrating metal atom}}

To determine the mean force worked on the migrating metal atom, first, the minimum-energy path and transition state for a single-vacancy hop in metals were determined using nudged elastic band (NEB) \cite{Henkelman2000}.
Next, the ensemble-averaged mean force $\langle \mathbf{F_X} \rangle$ was evaluated with the reaction coordinate $\lambda$ held fixed, where $\mathbf{F_X}$ is the force acting on the migrating metal atom and $\lambda$ was defined as the displacement of that atom relative to the center of mass of the remaining metal atoms.
A short-range repulsive pseudo-particle was placed at the vacancy center to prevent spontaneous collapse of nearby metal atoms into the vacancy during constrained sampling. For each fixed $\lambda$, $\mathbf{F_X}$ was estimated by averaging over \(8\times10^{5}\) MD steps following a \(1\times10^{5}\) steps equilibration.
For H-charged systems, H atoms were inserted/deleted using hybrid MD/GCMC at fixed hydrogen chemical potential.

\subsection*{\textbf{Direct molecular dynamics estimate from the Einstein relation (MDMSD)}}

In MD simulation, we compute the diffusion coefficient directly from the Einstein relation \cite{Maginn2018},
\begin{align}
	D_{\rm MD} = \lim_{t\to\infty}\frac{1}{2d}\frac{\dd}{\dd t}
	\left\langle \left|\mathbf R(t)-\mathbf R(0)\right|^2 \right\rangle,
	\label{eq:einstein}
\end{align}
where $\mathbf R(t)$ is the vacancy position and $d$ is the dimensionality. In this work we take $d=3$ for three dimensional vacancy diffusion.

The mean squared displacement was determined via MD simulations in the NPT ensemble by averaging over \(2\times10^{6}\) MD steps following a \(5\times10^{5}\) steps equilibration. Models with 1024 atoms for Fe, 2048 atoms for Cu, Pd, and Ni, 1440 atoms for Zr, and 3456 atoms for W were employed. A single monovacancy was created by removing one metal atom. We evaluate $D_{\rm vac}$ for both the charged system and the uncharged reference, and then report their ratio. For charged system, hydrogen atoms were introduced either using a hybrid MD/GCMC method at fixed hydrogen chemical potential $\mu$, or by simulations with a fixed number of hydrogen atoms.

\section*{Appendix}

\subsection*{\textbf{A. Grand canonical ensemble for mobile interstitials}}

We consider a crystalline host with a fixed number of lattice atoms $N_{\mathrm{lat}}$ at fixed $V$ and $T$, while the number of mobile interstitial atoms of a species $\I$ fluctuates under a reservoir chemical potential $\mu \equiv \mu_{\I}$. We use $\alpha$ to denote a generic external parameter in the Hamiltonian.

The grand partition function is
\begin{align}
	\Xi(\mu,V,T;\alpha)
	&=\sum_{N=0}^{\infty}
	\frac{z^N}{N!\,h^{3(N_{\mathrm{lat}}+N)}}
	\int \dd\mathbf X\,\dd\mathbf P
	\int \prod_{i=1}^{N}\dd\mathbf x_i\,\dd\mathbf p_i\;
	\e^{-\beta H(\mathbf X,\mathbf P,\{\mathbf x_i,\mathbf p_i\}_{i=1}^{N};N,\alpha)},
	\label{eq:Xi_def}
	\\
	z&\equiv \e^{\beta\mu},\qquad \beta\equiv (\kB T)^{-1}.
\end{align}
Here $N$ denotes the number of mobile interstitial atoms of species $\I$ in the system.
The grand potential is
\begin{align}
	\Omega(\alpha) = -\kB T \ln \Xi(\mu,V,T;\alpha).
	\label{eq:Omega_def}
\end{align}
Differentiating Eq. \eqref{eq:Omega_def} yields
\begin{align}
	\dv{\Omega}{\alpha}
	=
	\left\langle \pdv{H}{\alpha} \right\rangle_{\mu VT,\alpha},
	\qquad
	\Omega(\alpha_1)-\Omega(\alpha_0)
	=
	\int_{\alpha_0}^{\alpha_1}\left\langle \pdv{H}{\alpha} \right\rangle_{\mu VT,\alpha}\dd\alpha,
	\label{eq:dOmega_dalpha_explicit}
\end{align}
where $\langle \cdots \rangle_{\mu VT,\alpha}$ denotes a grand canonical average at fixed $\mu$, $V$, and $T$.

\subsection*{\textbf{B. Hyperplane constraints along a reference reaction pathway}}

Let the host lattice configuration vector be
\begin{align}
	\mathbf X \equiv (\mathbf R_1,\ldots,\mathbf R_{N_{\mathrm{lat}}})\in\mathbb R^{3N_{\mathrm{lat}}},
\end{align}
with global translations and rigid rotations removed by fixing the center of mass and the orientation of the host lattice configuration (for example by aligning principal axes or fixing three reference atoms).
We represent a reference defect migration pathway by a continuously differentiable curve
\begin{align}
	\bar{\mathbf X}(\Lambda)\in\mathbb R^{3N_{\mathrm{lat}}}, \ \Lambda\in[0,1],
\end{align}
where $\Lambda$ parameterizes the pathway. 

Without loss of generality, $\Lambda$ may be chosen as the normalized arc length along $\bar{\mathbf X}(\Lambda)$.
Define the unit tangent vector
\begin{align}
	\hat{\mathbf t}(\Lambda)\equiv
	\frac{\dv{\bar{\mathbf X}}{\Lambda}}{\left\|\dv{\bar{\mathbf X}}{\Lambda}\right\|}.
\end{align}
For each $\Lambda$, we define a $(3N_{\mathrm{lat}}-1)$ dimensional hyperplane orthogonal to the tangent as
\begin{align}
	\mathcal H_{\Lambda}\equiv
	\left\{\mathbf X:\ \hat{\mathbf t}(\Lambda)\cdot\big(\mathbf X-\bar{\mathbf X}(\Lambda)\big)=0\right\}.
	\label{eq:hyperplane_family_continuous}
\end{align}
The family $\{\mathcal H_{\Lambda}\}_{\Lambda\in[0,1]}$ provides a continuous set of constraints along the reference pathway.

In numerical calculations, the continuous reference pathway $\bar{\mathbf X}(\Lambda)$ is represented by a discrete set of images
$\{\mathbf X^{(k)}\}_{k=0}^{K}$ obtained from standard pathway optimization methods such as the nudged elastic band method \cite{Henkelman2000NEB} or the string method \cite{E2002String}, with $\mathbf X^{(0)}$ the initial configuration.
We then approximate $\bar{\mathbf X}(\Lambda)$ by piecewise linear segments connecting neighboring images.
Define segment vectors and lengths
\begin{eqnarray}
	&&\Delta\mathbf X^{(k)} \equiv \mathbf X^{(k+1)}-\mathbf X^{(k)}, \
	L_k \equiv \|\Delta\mathbf X^{(k)}\|, \nonumber \\
	&&\hat{\mathbf n}_k \equiv \frac{\Delta\mathbf X^{(k)}}{L_k}.
\end{eqnarray}
Introduce a global arc length parameter
\begin{align}
	S_0 \equiv 0,\
	S_{k+1}\equiv S_k+L_k,\
	S_{\mathrm{tot}}\equiv S_K=\sum_{k=0}^{K-1}L_k,
\end{align}
and the normalized pathway coordinate $\Lambda\in[0,1]$ by
\begin{align}
	\Lambda \equiv \frac{s}{S_{\mathrm{tot}}}, \ s\in[0,S_{\mathrm{tot}}].
\end{align}
For $\Lambda$ such that $s=\Lambda S_{\mathrm{tot}}\in[S_k,S_{k+1}]$, we introduce a local segment parameter
\begin{align}
	\lambda_k(\Lambda)\equiv \frac{s-S_k}{L_k}\in[0,1].
	\label{eq:local_lambda_def}
\end{align}
We then define, on each segment $k$, a family of linear hyperplanes
\begin{align}
	\mathcal H_{k,\lambda}\equiv
	\left\{\mathbf X:\ \hat{\mathbf n}_k\cdot(\mathbf X-\mathbf X^{(k)})=\lambda L_k\right\},
	\ \lambda\in[0,1].
	\label{eq:hyperplane_family_segment}
\end{align}

\subsection*{\textbf{C. Continuous formulation of the hyperplane-constrained grand potential}}

For the continuous hyperplane family $\{\mathcal H_{\Lambda}\}_{\Lambda\in[0,1]}$ defined in Eq.~(\ref{eq:hyperplane_family_continuous}), we define the hyperplane constrained grand partition function by restricting the host lattice configuration to $\mathcal H_{\Lambda}$:
\begin{equation}
	\begin{aligned}
		&\Xi_{\mathcal H}(\mu,V,T;\Lambda)=\\
		&\sum_{N=0}^{\infty}
		\frac{z^N}{N!\,h^{3(N_{\mathrm{lat}}+N)}}
		\int \dd\mathbf X\,\dd\mathbf P
		\int \prod_{i=1}^{N}\dd\mathbf x_i\,\dd\mathbf p_i\;
		\delta\!\Big(\hat{\mathbf t}(\Lambda)\cdot(\mathbf X-\bar{\mathbf X}(\Lambda))\Big)\,
		\e^{-\beta H(\mathbf X,\mathbf P,\{\mathbf x_i,\mathbf p_i\};N)}.
	\end{aligned}
	\label{eq:Xi_hyperplane_continuous}
\end{equation}
The corresponding constrained grand potential is
\begin{align}
	\Omega_{\mathcal H}(\Lambda;\mu,T)\equiv -\kB T\ln \Xi_{\mathcal H}(\mu,V,T;\Lambda),
	\label{eq:Omega_hyperplane_continuous}
\end{align}
and the pathway free energy profile is defined as
\begin{align}
	\Delta G(\Lambda;T,\mu)\equiv
	\Omega_{\mathcal H}(\Lambda;\mu,T)-\Omega_{\mathcal H}(0;\mu,T).
	\label{eq:DG_continuous_def}
\end{align}

\subsection*{\textbf{D. Relation of hyperplane-constrained grand potential to an equilibrium path ensemble and configuration conditioned profiles}}

Equation~(\ref{eq:Xi_hyperplane_segment}) defines the hyperplane constrained grand partition function
$\Xi_{\mathcal H}^{(k)}(\mu,V,T;\lambda)$ as a conditional functional integral at fixed $\mu$ that sums over the fluctuating interstitial number and integrates over all interstitial configurations compatible with the host constraint $\mathbf X\in\mathcal H_{k,\lambda}$.
Consequently, Eq. (\ref{eq:Omega_hyperplane_segment}) is the negative logarithm of this conditional integral, and Eq. (\ref{eq:DG_segment_def}) is a potential of mean force for the slow defect coordinate on segment $k$ that already incorporates the statistical weight of many microscopic interstitial rearrangements.
In this sense, the present framework can be viewed as an equilibrium reduction of a path ensemble in which repeated activated events sample different instantaneous interstitial configurations at fixed $\mu$ (Fig.~\ref{fig:schematic_TI}).

To connect this formal definition to the intuitive picture, it is useful to explicitly separate contributions from interstitial realizations.
Let $\nu$ label a microscopic interstitial realization compatible with the hyperplane constraint $\mathcal H_{k,\lambda}$, including the interstitial number $N_\nu$ and the set of interstitial coordinates (or site occupancies) in that realization.
Here $\{\mathbf x\}_{\nu}$ denotes the collection of interstitial coordinates (or discrete site occupancies) in realization $\nu$, sampled from the hyperplane constrained grand canonical distribution at fixed $(\mu,V,T)$.
After integrating out momenta, each realization contributes a positive configurational weight
\begin{align}
	w_{\nu}^{(k)}(\lambda;\mu,T)\equiv \exp\!\big(\beta\mu N_{\nu}\big)\,
	\int_{\mathcal H_{k,\lambda}} \dd \mathbf X \;
	\exp\!\big[-\beta U(\mathbf X,\{\mathbf x\}_{\nu})\big],
	\label{eq:wi_def}
\end{align}
where the host integration is restricted to $\mathcal H_{k,\lambda}$ and $U$ is the potential energy.
We absorb factors that are independent of $\lambda$ into the definition of $w_{\nu}^{(k)}$, since they cancel in free energy differences.
Defining the corresponding realization conditioned free energy
\begin{align}
	\mathcal G_{\nu}^{(k)}(\lambda;T,\mu)\equiv -\kB T\ln w_{\nu}^{(k)}(\lambda;\mu,T),
	\label{eq:Gi_def}
\end{align}
the constrained grand partition function can be written as $\Xi_{\mathcal H}^{(k)}(\lambda)=\sum_{\nu} w_{\nu}^{(k)}(\lambda)$, and therefore the segment free energy profile can be expressed as
\begin{eqnarray}
	&&\Delta G_k(\lambda;T,\mu) \nonumber \\
	&&=
	-\kB T \ln\!\left(
	\frac{\sum_{\nu} \exp\!\big[-\beta \mathcal G_{\nu}^{(k)}(\lambda;T,\mu)\big]}
	{\sum_{\nu} \exp\!\big[-\beta \mathcal G_{\nu}^{(k)}(0;T,\mu)\big]}
	\right).
	\label{eq:DG_logsum_ratio}
\end{eqnarray}
Equivalently, introducing $\Delta G_{\nu}^{(k)}(\lambda)\equiv \mathcal G_{\nu}^{(k)}(\lambda)-\mathcal G_{\nu}^{(k)}(0)$, one may view $\Delta G_k(\lambda;T,\mu)$ as the statistical envelope of a family of realization dependent profiles $\Delta G_{\nu}^{(k)}(\lambda)$ corresponding to different interstitial arrangements across repeated activated events.
The global profile $\Delta G(\Lambda;T,\mu)$ obtained by concatenation in the piecewise linear representation: for $s=\Lambda S_{\mathrm{tot}}\in[S_k,S_{k+1}]$, $\Delta G(\Lambda;T,\mu) \approx \sum_{m=0}^{k-1}\Delta G_m(1;T,\mu) + \Delta G_k(\lambda_k(\Lambda);T,\mu)$, inherits the same interpretation at each constrained state along the pathway. When interstitial redistribution remains fast enough to maintain quasi equilibrium with the reservoir during repeated events, this equilibrium marginalization provides the thermodynamic input consistent with ensemble averaged kinetics. When strong driving forces cause the interstitial atmosphere to lag behind the moving defect, nonequilibrium trajectory level effects such as drag and memory can become important and are beyond the present equilibrium activation free energy description.

\subsection*{\textbf{E. Generalized mean force for a moving hyperplane constraint}}

For the continuous hyperplane family $\{\mathcal H_{\Lambda}\}$ defined by Eq.~(\ref{eq:hyperplane_family_continuous}), it is convenient to introduce the constraint function
\begin{align}
	\sigma(\mathbf X;\Lambda)\equiv \hat{\mathbf t}(\Lambda)\cdot\big(\mathbf X-\bar{\mathbf X}(\Lambda)\big)=0,
	\label{eq:sigma_continuous}
\end{align}
which specifies $\mathbf X\in\mathcal H_{\Lambda}$.
In constrained sampling, the derivative of the constrained grand potential with respect to the pathway coordinate can be written as an ensemble average of the generalized force conjugate to $\Lambda$,
\begin{align}
	\dv{\Omega_{\mathcal H}(\Lambda;\mu,T)}{\Lambda}
	=
	\left\langle \zeta_{\Lambda}\,\pdv{\sigma(\mathbf X;\Lambda)}{\Lambda}\right\rangle_{\mu VT,\Lambda},
	\label{eq:dOmega_dLambda_general}
\end{align}
where $\zeta_{\Lambda}$ is the Lagrange multiplier associated with enforcing $\sigma(\mathbf X;\Lambda)=0$.
The derivative of the constraint contains a translation term and a geometric term associated with the rotation of the hyperplane along a curved pathway,
\begin{align}
	\pdv{\sigma(\mathbf X;\Lambda)}{\Lambda}
	=
	\dv{\hat{\mathbf t}}{\Lambda}\cdot\big(\mathbf X-\bar{\mathbf X}(\Lambda)\big)
	-
	\hat{\mathbf t}(\Lambda)\cdot\dv{\bar{\mathbf X}}{\Lambda}.
	\label{eq:dSigma_dLambda}
\end{align}
In practice, we evaluate $\Delta G(\Lambda;T,\mu)$ by discretizing $\bar{\mathbf X}(\Lambda)$ into locally linear segments (piecewise linear approximation).
Within each segment, the tangent (and hence the hyperplane normal) is constant, so the geometric term proportional to $\dv{\hat{\mathbf t}}{\Lambda}$ vanishes.
This reduces the generalized mean force to the projected mean force for a fixed normal, leading to the segment based mean force thermodynamic integration implemented below.
As the maximum segment length is reduced (equivalently, as the number of images along the pathway is increased), the piecewise linear construction converges to the underlying continuous curve $\bar{\mathbf X}(\Lambda)$, and the resulting segment based thermodynamic integration approaches the continuous free energy profile $\Delta G(\Lambda;T,\mu)$ defined by Eqs.~(\ref{eq:Omega_hyperplane_continuous}) and (\ref{eq:DG_continuous_def}).

\section*{Data availability}
All the files used and/or constructed during the current study are available from the corresponding author on reasonable request.

\section*{Acknowledgements}
S.H.Zhang was funded by the JSPS KAKENHI Grant No. JP25K17509. S.O. acknowledges the support by the Ministry of Education, Culture, Sport, Science and Technology of Japan (Grant Nos. JPMXP1122684766, JPMXP1020230325, and JPMXP1020230327), and the support by JSPS KAKENHI (Grant Nos. JP23H00161 and JP23K20037). Part of the calculations were performed on the large-scale computer systems at the Cybermedia Center, The University of Osaka, the Large-scale parallel computing server at the Center for Computational Materials Science, Institute for Materials Research, Tohoku University, and supercomputer Fugaku provided by the RIKEN Center for Computational Science (Project IDs: hp250229 and hp250227). J.L. acknowledges support from the US DOE ARPA-E.

\section*{Contributions}

S. Zhang and S. Zhu: methodology, software, formal analysis, and original draft writing; J.D. and S.S.: writing—review \& editing; J.L.: conceptualization, writing—review \& editing, and supervision; S.O.: conceptualization, writing—review \& editing, supervision, and project management. All authors contributed to discussions and the final manuscript.

\section*{Competing interests}

The authors declare no competing interests.


\end{document}